\documentclass[12pt]{elsarticle}

\usepackage{amssymb}

\usepackage{lineno}

\usepackage{siunitx}
\usepackage{graphicx}
\usepackage{float}
\DeclareGraphicsExtensions{.png,.pdf,.jpg}

\usepackage[automake,abbreviations,symbols,nomain,toc,nogroupskip,nonumberlist]{glossaries-extra}
\GlsXtrEnableEntryCounting
 {abbreviation}
 {1}

\newabbreviation{aps}{APS}{autonomous power systems}
\newabbreviation{capex}{CAPEX}{capital expenditure}
\newabbreviation{ems}{EMS}{energy management system}
\newabbreviation{ess}{ESS}{energy storage system}
\newabbreviation{fcr}{FCR}{frequency containment reserves}
\newabbreviation{ffr}{FFR}{fast frequency reserves}
\newabbreviation{frr}{FRR}{frequency restoration reserves}
\newabbreviation{ghg}{GHG}{greenhouse gases}
\newabbreviation{gt}{GT}{gas turbine}
\newabbreviation{lcoe}{LCOE}{levelized cost of energy}
\newabbreviation{milp}{MILP}{mixed integer linear programming}
\newabbreviation{og}{O\&G}{oil and gas}
\newabbreviation{pu}{pu}{per unit}
\newabbreviation{pv}{PV}{photovoltaic}
\newabbreviation{res}{RES}{renewable energy source}
\newabbreviation{uc}{UC}{unit commitment}

\newignoredglossary{ignore}

\glsxtrnewsymbol[description={frequency deviation from its rated value},sort=fd]
    {df}
    {\ensuremath{\tilde{f}}}

\glsxtrnewsymbol[description={angular speed},sort=omega]
    {omega}
    {\ensuremath{\omega}}

\glsxtrnewsymbol[description={synchronous or rated angular speed},sort=omegas]
    {omegas}
    {\ensuremath{\omega_s}}

\glsxtrnewsymbol[description={total active power generation in Watts},sort=PG]
    {PG}
    {\ensuremath{P_G}}

\glsxtrnewsymbol[description={total active power consumption in Watts},sort=PL]
    {PL}
    {\ensuremath{P_L}}

\glsxtrnewsymbol[description={total power generation in pu},sort=pg]
    {pg}
    {\ensuremath{p_G}}

\glsxtrnewsymbol[description={total power consumption in pu},sort=pl]
    {pl}
    {\ensuremath{p_L}}

\glsxtrnewsymbol[description={total power generation at time step h},sort=pg]
    {pgh}
    {\ensuremath{p_{G,h}}}

\glsxtrnewsymbol[description={total power consumption at time step h},sort=plh]
    {plh}
    {\ensuremath{p_{L,h}}}

\glsxtrnewsymbol[description={equivalent moment of inertia},sort=J]
    {J}
    {\ensuremath{J}}

\glsxtrnewsymbol[description={equivalent damping},sort=B]
    {B}
    {\ensuremath{B}}
    
\glsxtrnewsymbol[description={generator torque},sort=TG]
    {TG}
    {\ensuremath{T_G}}

\glsxtrnewsymbol[description={load torque},sort=TL]
    {TL}
    {\ensuremath{T_L}}

\glsxtrnewsymbol[description={boundary for frequency deviation in steady-state},sort=rss]
    {rss}
    {\ensuremath{r_{ss}}}

\glsxtrnewsymbol[description={boundary for frequency deviation during large transients},sort=rtr]
    {rtr}
    {\ensuremath{r_{tr}}}

\glsxtrnewsymbol[description={normalized inertia},sort=M]
    {M}
    {\ensuremath{M}}

\glsxtrnewsymbol[description={normalized damping},sort=D]
    {D}
    {\ensuremath{D}}

\glsxtrnewsymbol[description={minimum controlled damping},sort=Dmin]
    {Dmin}
    {\ensuremath{D_{min}}}

\glsxtrnewsymbol[description={boundary for power imbalance},sort=Pb]
    {Pb}
    {\ensuremath{P_{b}}}

\glsxtrnewsymbol[description={boundary for power imbalance at time step h},sort=Pbh]
    {Pbh}
    {\ensuremath{P_{b,h}}}

\glsxtrnewsymbol[description={allocated FCR for generator m at time step h},sort=PFCRmh]
    {PFCRmh}
    {\ensuremath{P_{m,h}^{FCR}}}

\glsxtrnewsymbol[description={dispatch for generator m at time step h},sort=Pmh]
    {Pmh}
    {\ensuremath{P_{m,h}}}

\glsxtrnewsymbol[description={minimum dispatch for generator m},sort=Pminm]
    {Pminm}
    {\ensuremath{P_{m}^{min}}}

\glsxtrnewsymbol[description={maximum dispatch for generator m},sort=Pmaxm]
    {Pmaxm}
    {\ensuremath{P_{m}^{max}}}

\glsxtrnewsymbol[description={ramp rate of RES variation},sort=rrRES]
    {rrRES}
    {\ensuremath{rr^{RES}}}

\glsxtrnewsymbol[description={ramp rate of RES variation at time h},sort=rrRESh]
    {rrRESh}
    {\ensuremath{rr_{h}^{RES}}}

\glsxtrnewsymbol[description={ramp rate of FRR for generator m},sort=rrFFRm]
    {rrFFRm}
    {\ensuremath{rr_{m}^{FFR}}}

\glsxtrnewsymbol[description={duration of RES variation},sort=TRES]
    {TRES}
    {\ensuremath{\Delta T^{RES}}}

\glsxtrnewsymbol[description={duration of RES variation at time h},sort=TRESh]
    {TRESh}
    {\ensuremath{\Delta T_{h}^{RES}}}

\glsxtrnewsymbol[description={duration of ramp r at time h},sort=Trh]
    {Trh}
    {\ensuremath{\Delta T_{r,h}}}

\glsxtrnewsymbol[description={cost of installed solar PV capacity},sort=cPV]
    {cPV}
    {\ensuremath{c_{PV}}}

\glsxtrnewsymbol[description={installed solar PV capacity},sort=PPVinst]
    {PPVinst}
    {\ensuremath{P_{PV}^{inst}}}

\glsxtrnewsymbol[description={installed battery capacity},sort=Pbatinst]
    {Pbatinst}
    {\ensuremath{P_{bat}^{inst}}}

\glsxtrnewsymbol[description={required battery capacity for FCR at time step h},sort=PFCRbath]
    {PFCRbath}
    {\ensuremath{P_{bat,h}^{FCR}}}

\glsxtrnewsymbol[description={cost of installed battery capacity},sort=cbat]
    {cbat}
    {\ensuremath{c_{bat}}}

\glsxtrnewsymbol[description={fuel cost for generator m at time step h},sort=FCmh]
    {FCmh}
    {\ensuremath{FC_{m,h}}}

\glsxtrnewsymbol[description={cost of fuel},sort=cf]
    {cf}
    {\ensuremath{c_{f}}}

\glsxtrnewsymbol[description={variable fuel consumption for generator m},sort=am]
    {am}
    {\ensuremath{a_{m}}}

\glsxtrnewsymbol[description={fixed fuel consumption for generator m},sort=bm]
    {bm}
    {\ensuremath{b_{m}}}

\glsxtrnewsymbol[description={injected solar PV power at time step h},sort=PPVinjh]
    {PPVinjh}
    {\ensuremath{P_{PV,h}^{inj}}}

\glsxtrnewsymbol[description={available solar PV area at time step h},sort=APVh]
    {APVh}
    {\ensuremath{A_{PV,h}}}

\glsxtrnewsymbol[description={solar irradiance at time step h},sort=Ih]
    {Ih}
    {\ensuremath{I_{h}}}

\glsxtrnewsymbol[description={solar irradiance drop for ramp r at time step h},sort=Irh]
    {Irh}
    {\ensuremath{\Delta I_{r,h}}}

\glsxtrnewsymbol[description={on/off status of generator m at time step h},sort=rhomh]
    {rhomh}
    {\ensuremath{\rho_{m,h}}}

\glsxtrnewsymbol[description={start-up decision of generator m at time step h},sort=umh]
    {umh}
    {\ensuremath{u_{m,h}}}

\glsxtrnewsymbol[description={shut-down decision of generator m at time step h},sort=vmh]
    {vmh}
    {\ensuremath{v_{m,h}}}

\glsxtrnewsymbol[description={minimum down-time of generator m},sort=Tmdn]
    {Tmdn}
    {\ensuremath{T_{m}^{dn}}}

\glsxtrnewsymbol[description={minimum up-time of generator m},sort=Tmup]
    {Tmup}
    {\ensuremath{T_{m}^{up}}}

\glsxtrnewsymbol[description={boundary for power imbalance due to solar PV short-term variation at time step h},sort=PbhPV]
    {PbhPV}
    {\ensuremath{P_{b,h}^{PV}}}

\glsxtrnewsymbol[description={boundary for power imbalance due to sudden load variation or generation loss at time step h},sort=Pbhsud]
    {Pbhsud}
    {\ensuremath{P_{b,h}^{sud}}}
    
\glsxtrnewsymbol[description={solar PV power drop for ramp r at time step h},sort=PPVrh]
    {PPVrh}
    {\ensuremath{\Delta P_{r,h}^{PV}}}

\glsxtrnewsymbol[description={total apparent power of generators},sort=Sn]
    {Sn}
    {\ensuremath{S_{n}}}

\glsxtrnewsymbol[description={discount rate of the project},sort=e]
    {e}
    {\ensuremath{e}}
    
\glsxtrnewsymbol[description={P-f droop of generator m},sort=betam]
    {betam}
    {\ensuremath{\beta_{m}}}

\makeglossaries

\journal{Applied Energy}

\usepackage{cleveref}

\begin{document}
\glsaddall

\begin{frontmatter}





\title{Allocation of spinning reserves for autonomous grids subject to frequency stability constraints and short-term solar power variations}



\author[inst3]{Erick Fernando Alves} 
\author[inst1,inst2]{Louis Polleux}
\author[inst1]{Gilles Guerassimoff}
\author[inst3]{Magnus Korp{\aa}s} 
\author[inst3,inst4]{Elisabetta Tedeschi}

\affiliation[inst3]{organization={Department of Electric Power Engineering, Norwegian University of Science and Technology},
            addressline={O.S. Bragstads Plass 2E}, 
            city={Trondheim},
            postcode={7034}, 
            country={Norway}}

\affiliation[inst1]{organization={Center for Applied Mathematics, Mines Paristech PSL},
            addressline={1 Rue Claude Daunesse}, 
            city={Sophia Antipolis},
            postcode={06904}, 
            country={France}}

\affiliation[inst2]{organization={Research and Development, TotalEnergies},
            addressline={2 Place Jean Millier}, 
            city={Courbevoie},
            postcode={92400}, 
            country={France}}

\affiliation[inst4]{organization={
Department of Industrial Engineering, University of Trento},
            addressline={Via Sommarive, 9}, 
            city={Povo},
            postcode={38123}, 
            state={Trento},
            country={Italy}}

\begin{abstract}
Low-inertia, isolated power systems face the problem of resiliency to power variations.
The integration of renewable energy sources, such as wind and solar \glsxtrlong{pv}, pushes the boundaries of this issue further.
Higher shares of renewables requires better evaluations of electrical system stability, to avoid severe safety and economic consequences.
Accounting for frequency stability requirements and allocating proper spinning reserves, therefore becomes a topic of pivotal importance in the long-term planning and operational management of power systems.
In this paper, dynamic frequency constraints are proposed to ensure resiliency during short-term power variations due to, for example, wind gusts or cloud passage.
The use of the proposed constraints is exemplified in a case study, the constraints being integrated into a mixed-integer linear programming algorithm for sizing the optimal capacities of solar \glsxtrlong{pv} and battery energy storage resources in an isolated industrial plant.
Outcomes of this case study show that reductions in the levelized cost of energy and carbon emissions can be overestimated by 8.0\% and 10.8\% respectively, where frequency constraints are neglected.
The proposed optimal sizing is validated using time-domain simulations of the case study.
The results indicate that this optimal system is frequency stable under the worst-case contingency.
\end{abstract}


\begin{highlights}
\item We propose constraints for contingency and intermittence resilience in renewable power
\item We integrate frequency constraints in a sizing problem, through using MILP
\item We consider limited ramp capacity and different frequency control schemes
\item We use solar ramp worst-case scenarios to calculate primary storage requirements 
\end{highlights}


\end{frontmatter}


\section{Glossary}
\printglossary[type=abbreviations]
\printglossary[type=symbols]

\section{Introduction}
\label{sec:intro}

The depletion of mature \gls{og} fields reduces the energy return on investment of the field and increases the emission of nitrogen oxides and \gls{ghg} \cite{masnadiClimateImpactsOil2017}.
The use of \glspl{res} in \gls{og} operations has therefore become an active research area in recent years.
Groups have analyzed this challenge from a broad range of perspectives \cite{riboldiIntegratedAssessmentEnvironmental2019, polleuxImpactsThermalGeneration2019, itikiComprehensiveReviewProposed2019, chapaloglouTechnoEconomicEvaluationSizing2019, riboldiOptimalDesignHybrid2020, teeTransientStabilityAnalysis2020, anglaniRenewableEnergySources2020,khanEvaluationDeepWaterFloating2021}.
It is, however, widely accepted that the planning or operation of hybrid energy systems in the \gls{og} industry requires the solution of an optimization problem.
The objective of such optimization can differ considerably.
The power balance of the energy system is, however, a physical constraint that must always be satisfied.

Most \gls{og} plants, platforms and vessels are supplied ac power by low-inertia, isolated systems \cite{devoldOilGasProduction2013}.
Large active power variations in and considerable deviations from the system rated frequency can, however, occur where the penetration of \glspl{res} in such systems is high.
This is a problem that is also found in other \gls{aps}, such as islands \cite{vasconcelosAdvancedControlSolutions2015,ringkjobTransitioningRemoteArctic2020} and community microgrids \cite{meenaOptimisationFrameworkDesign2019}.
Frequency stability constraints and the proper sizing of spinning reserves, such as \gls{fcr} and \gls{frr}, are therefore important dimensioning aspects and should therefore be given special attention in optimization algorithms \cite{rebollal_endogenous_2021}.

Short-term variations caused by cloud passage across solar \gls{pv} parks or wind gusts at wind farms can furthermore saturate the ramping capabilities of other generators.
The relationship between \glspl{res} penetration and variability, size of \glspl{ess}, scheduling decisions and frequency stability should therefore be investigated in more detail.
Large interconnected and meshed electricity networks also are progressively behaving more as low-inertia systems, as the green shift progresses worldwide and penetration of \glspl{res} continues \cite{orumFutureSystemInertia2018,holttinenSystemImpactStudies2020a}.
Interest in the topic of this paper therefore is wide.

\subsection{Literature review}
\label{sec:intro-review}
Reliability-constrained \gls{uc} is described, for example, in \cite{fotouhi_ghazvini_coordination_2012,goleijaniReliabilityConstrainedUnit2013,mashayekh_integrated_2015},
and addresses the \gls{uc} problem during unexpected events such as the severe loss of load, generation or transmission capacity.
The strategies applied to mitigate this include n+1 redundancy, and spinning reserve for frequency containment and restoration.
\cite{rezaei_economicenvironmental_2014} for example proposed a two-stage stochastic \gls{milp} formulation for the calculation of  \gls{uc} and reserve scheduling, at frequency deviations of 0 and \SI{\pm10}{\milli\hertz}.
\cite{wenFrequencyDynamicsConstrained2016, wen_frequency_2016-1} also discuss formulations for addressing wind power variations and the correct allocation from \glspl{ess} of spinning reserves.
Generator contingency was, in this, considered to be the largest wind power fluctuation from one optimization time-step to another.
Power balances were handled before and after contingencies in these works, and without generator transient period dynamics or ramping capacities being taken into account.

\cite{cardozoFrequencyConstrainedUnit2017} have proposed frequency-constrained \gls{uc} problem implementation alternatives to circumvent these limitations.
\cite{arteaga_optimisation_2016,yin_frequency-constrained_2021} thoroughly discuss the challenges of a frequency-constrained model.
Options such as the use of maximum ramping capacity or minimum value of inertia were evaluated, but can lead to a nonlinear problem.
This can, however, be handled by a Benders decomposition.
\cite{nguyenOptimalPowerFlow2019} introduced a nonlinear model that includes frequency security constraints and the dynamics of the transient period.
This, however, requires the use of a genetic algorithm to solve the optimization problem.
\cite{riboldiOptimalDesignHybrid2020} also use a genetic algorithm to optimally size the energy system of an offshore platform interconnected with a wind farm. 
The maximum ramping capacity of generators and frequency stability limits were furthermore evaluated using time-domain simulations of worst-case scenarios for each candidate solution.

\cite{schittekatte_impact_2018} integrated the power drops observed in solar \gls{pv}, within 15-minute time windows, into the constraints of an optimal planning problem, this accounting for the stochastic variation of solar irradiance.
A recent work \cite{polleuxRelationshipBatteryPower2021}, however, highlights the varying duration and magnitude of the short-term variability of \gls{pv} plants, and also that maximum perturbation may rise to \SI{60}{\percent} of the plant's rated power in less than \SI{30}{\second}.
These short-term variations must be taken into account if realistic decisions on storage investment or unit scheduling in optimization algorithms are to be achieved.
\cite{rebollal_endogenous_2021} furthermore provide an updated review of the frequency-constrained \gls{uc} problem, a classification of the different approaches documented in the literature, and a discussion of their many shortcomings.
They also proposed a linear model that includes frequency constraints directly in a \gls{milp} algorithm, and that therefore does not require the use of external time-domain simulations in the assessment of solution security.
They did not, however, consider the effect of fast short-term variations in \glspl{res}, combined with generator ramping restrictions and use of \glspl{ess} on frequency stability.

Challenges therefore still remain in the efficient sizing, planning and operation of autonomous, low-inertia power systems with high  \glspl{res} penetration, especially wind and solar \gls{pv}.

\subsection{Paper contributions}
\label{sec:intro-contrib}
This paper addresses some of the modeling challenges and shortcomings described above,
and introduces a set of algebraic frequency stability constraints that can be directly applied to linear optimization formulations of the \gls{uc} problem in low-inertia \gls{aps} with high \glspl{res} penetration.
Practical issues such as limited generator ramping capacities and variability of \glspl{res} are taken into consideration.
Short-term frequency variations are also used to reduce the conservatism of a \gls{uc} under uncertainty, which permits complementary \gls{fcr} and \gls{frr} action and a reduction in the power required from \glspl{ess}.

These ideas are exemplified in a case study of the integration of a solar \gls{pv} plant and a battery \gls{ess} into an existing isolated industrial installation fed by \glspl{gt}.
A \gls{milp} algorithm is used to solve a rolling frequency-constrained \gls{uc} problem, and to optimally size the installed capacity of solar \gls{pv} and battery \gls{ess}.
The \gls{res} short-term variability is evaluated by using a method recently published in \cite{polleuxRelationshipBatteryPower2021}, that identifies equivalent solar ramping scenarios during cloud passage.
The novel proposal, described in detail in \cref{sec:form2}, are linear constraints for frequency stability that model tolerated frequency dynamics and the complementary \gls{fcr} and \gls{frr} action taking place during short-term power variations, such as cloud passage events in solar \gls{pv} installations or wind gusts in wind farms.
This allows a significant reduction in the amount of \gls{fcr} and consequently the rated power of an \gls{ess} designed to support the grid in such events, as demonstrated by the results of the case study in \cref{sec:case-comp}.
The frequency-stability results for the constrained and non-constrained implementation were compared, and major differences were highlighted, being the solution data obtained from this also used in a time-domain simulation to validate the results under the worst-case scenario.

\section{Addressing frequency stability with linear constraints}
This section reviews the equations that describe the frequency dynamics of a power system, and the types of spinning reserves required to ensure frequency stability.
The rationale for determining spinning reserves by obtaining a set of algebraic expressions from the dynamic equations, and applying these as constraints to the linear optimization problem, is also introduced.
The first proposed formulation treats the sudden disconnection of loads or generators.
The next formulation addresses the problem of the short-term variation of loads or \glspl{res}, this requiring gradual compensation by dispatchable generators with limited ramping capacity.

\subsection{Frequency stability and spinning reserves}\label{sec:freqstab}
The simplified model of a flywheel spinning at angular speed \gls{omegas} can be used to express the dynamics of frequency in a power system.
The rotating masses of all synchronous generators and motors are represented by an equivalent moment of inertia $J$.
Generators deliver, at one end of the shaft, energy to the flywheel via torque $T_G$. Load removes energy through $T_L$ at the other shaft end.
The natural and controlled damping of the system is represented by coefficient $B$.
Applying Newton's second law of motion to this simplified model, and multiplying both sides by the angular speed \gls{omega}, therefore gives the following equation \cite{alvesSizingHybridEnergy2021}:
\begin{align}\label{eq:rotMass}
	\gls{omega} J \dot{\gls{omega}} = \gls{PG} - \gls{PL} - B(\gls{omega} - \gls{omegas})
\end{align}
where \gls{PG}, \gls{PL} are respectively the active power delivered by generators and consumed by loads.

Eq. \ref{eq:rotMass}, which is also known as the Swing Equation, represents the power balance that is required to maintain this system at its rated angular speed \gls{omegas}.
The frequency deviation $\gls{df}=\frac{\gls{omega}-\gls{omegas}}{\gls{omegas}}$ can, after this equation has been normalized by \gls{omegas} and by the total apparent power of the generators $S_n$, be estimated by \cite{alvesSizingHybridEnergy2021}:
\begin{align}\label{eq:deltaf}
	(1 + \gls{df}) M \dot{\gls{df}} = \gls{pg} - \gls{pl} - D \gls{df}
\end{align}
where $M$, $D$, \gls{pg}, \gls{pl} are respectively the normalized inertia, damping, and active powers delivered by generators and consumed by loads.

The system is considered to be frequency stable \cite{hatziargyriouDefinitionClassificationPower2021} where any power imbalance \gls{df} remains within intervals $\pm r_{tr}$ during transient conditions and $\pm r_{ss}$ during steady-state conditions.
These intervals are often defined in grid codes or industry standards.
In Europe, $r_{ss}=\pm 1\%$ and $r_{tr}=\pm 3\%$ are specified for generators connected to transmission \cite{comissionregulationeuNetworkCodeRequirements2016} or distribution systems \cite{NEK50549120192019,NEK50549220192019}.
Wider limits are normally specified for low-inertia, autonomous systems, for example
$r_{ss}$ between 2 and 5\%, and $r_{tr}$ between 5 and 10\% for fixed and mobile offshore units \cite{IEC6189220192019}.

The minimum controlled damping $D_{min}$ that generators must provide when assuming a steady-state in Eq.\ref{eq:deltaf} ($\dot{\gls{df}}=0$), and $\gls{df}<1\%$ and null natural damping from loads, can be approximated by:
\begin{align}\label{eq:dmin}
	D_{min} r_{ss} \geq P_{b}
\end{align}
where $P_{b}=(\gls{pg} - \gls{pl})$ represents the maximum continuous imbalance that the power system can withstand and still maintain frequency stability.
$P_{b}$ also defines the minimum level of spinning reserves that are required for frequency variations to be kept within the permitted range of $\pm r_{ss}$.
This is therefore referred to as \glsxtrfull{fcr} \cite{comissionregulationeuGuidelineElectricityTransmission2017}.
The assumptions used to derive Eq. \ref{eq:dmin} may underestimate $D_{min}$ when \gls{df} values of more than 1\% are possible  \cite{alvesSufficientConditionsRobust2020a}.
A more robust approximation must therefore be given:
\begin{align}\label{eq:dminrob}
	D_{min} r_{ss}(1-r_{tr}) \geq P_{b}
\end{align}

Note that the \gls{fcr} strategy (proportional control) cannot restore frequency to its rated value.
Achieving the zero steady-state error ($\gls{df}=0$) therefore requires \gls{pg} and \gls{pl} to be matched by generator re-dispatching, through integral control \cite{sauerPowerSystemDynamics2017,alvesSizingHybridEnergy2021}.
This requires additional spinning reserves, which are referred to as \glsxtrfull{frr} \cite{comissionregulationeuGuidelineElectricityTransmission2017}.
The action of \gls{frr}, \gls{fcr} constraining frequency during disturbances to defined bands, therefore can be slow.
This is desirable, not just to avoid new disturbances, but also to accommodate the ramping limitations of generators.

A single generator can provide both \gls{fcr} and \gls{frr}.
This may not, however, be the optimal solution from a technical or an economic perspective.
\Glspl{gt} can, for example, usually provide both types of reserves simultaneously.
Excessive activation of \gls{fcr} on \glspl{gt} can, however, increase actuator wear and tear and increase maintenance costs.
A more suitable \gls{fcr} could be batteries designed to supply large amounts of power for short periods of time.
Their participation in \gls{frr} will, however, usually require large energy storage capacity, which will increase investment costs and space requirements.

The optimal allocation of spinning reserves can, even for small power systems, therefore quickly become a complex process.
An optimization algorithm is therefore often used in sizing and operational decisions. 
The following sections present the proposed algebraic constraints.
These are based on the frequency dynamics of a power system, and can help in the determination of the optimal allocation of \gls{fcr} and \gls{frr}.

\subsection{Formulation 1: constraints for sudden load or generation loss}\label{sec:form1}

The following constraints must be met when considering the assumptions and conditions imposed by Eq. \ref{eq:dmin}, for generators participating in \gls{fcr}:
\begin{align} 
    &\forall m, \; \forall h, &P_{m,h}^{FCR} &\leq D_{m}r_{ss} \label{eq:PmhFCR} \\
	&\forall m, \; \forall h, &P_{m}^{min} + P_{m,h}^{FCR} &\leq P_{m,h} \leq P_{m}^{max} - P_{m,h}^{FCR} \label{eq:Pmhdroop}
\end{align}
where $P_{m,h}^{FCR}, P_{m,h}^{min}, P_{m,h}^{max}, P_{m,h}$ are respectively the assigned \gls{fcr} contribution, the minimum, the maximum and the current commitment of generator $m$ in time step $h$.

The sum of \gls{fcr} provided by all generators is to also, at each time step $h$, be greater than the worst-case power disturbance $P_{b,h}$:
\begin{align}
	&\forall h,\quad \sum_{m} P_{m,h}^{FCR} \geq P_{b,h} \label{eq:sumDDmin}
\end{align}

$P_{b,h}$ varies with time, because it depends on the worst expected sudden load or generation loss for the current state of the power system.
This value is normally, in a security assessment, obtained from the evaluation of possible contingencies.
This complex topic is, however, beyond the scope of this work, a simplified approach however being presented in section \ref{sec:sizereserves} and used in this paper.
A deeper discussion of this can be found in \cite{ciapessoniProbabilisticRiskBasedSecurity2016}.
Note that Eq. \ref{eq:PmhFCR} can be rewritten based on the assumptions and conditions imposed by Eq. \ref{eq:dminrob}, where  \gls{df} values of more than 1\% are permitted.

\subsection{Formulation 2: constraints for short-term power variations}
\label{sec:form2} 

In this paper, short-term power variations are considered to be variations that can be compensated for by generators that provide \gls{frr} without infringing their permitted ramping rates, and by generators that provide \gls{fcr} without infringing the permitted frequency variations in steady-state.
Examples of these variations in plants fed by \gls{res} include cloud passage across solar \gls{pv} parks or wind gusts at wind farms.

This formulation requires the following assumptions:
\begin{enumerate}
    \item A power system in balance and in steady-state before \gls{res} variations at ${t=t_0}$. In other words, in Eq.\ref{eq:deltaf} ${\gls{df}(t_0) = 0}, {\dot{\gls{df}}(t_0) = 0}$ and ${\gls{pg}(t_0)=\gls{pl}(t_0)}$. \label{as:ss}

    \item Worst-case short-term variations can be approximated by a maximum amplitude $\Delta p^{RES}$, a minimum duration $\Delta T^{RES}$, and a constant rate of change ${rr^{RES}=\Delta p^{RES} / \Delta T^{RES}}$.
    This assumption has been corroborated by recently published analysis of short-term wind  \cite{hannesdottirDetectionCharacterizationExtreme2019} and solar \cite{polleuxRelationshipBatteryPower2021} variability,
     the determination of these parameters from high-resolution wind speed and solar irradiance measurements also being discussed in detail in these papers.\label{as:rrRES}
    
    \item $\Delta p^{RES}$ is compensated for without infringing the maximum ramping rate $rr^{FRR}$ of the generators participating in \gls{frr}.\label{as:rrFRR}
    
    \item Short-term \gls{res} power variations are sufficiently smooth.
    It is therefore reasonable to assume minimal influence of system inertia ($M{\dot{\gls{df}} \approx 0}$) during interval $\Delta T^{RES}$.\label{as:ddf0}
\end{enumerate}

Assumption \ref{as:ss} implies that any power variation in the system can be arbitrarily attributed to either \gls{pl} or \gls{pg} in Eq.~\ref{eq:deltaf}.
Eq. \ref{eq:pl} is obtained based on the assumption that \gls{res} variations affect \gls{pl} and on assumption \ref{as:rrRES}.
Eq. \ref{eq:pg} also accounts for the contributions of all generators that provide \gls{frr} in term \gls{pg}, on assumption \ref{as:rrFRR} being here applied.
\begin{align}
	&\forall h, &p_{L,h} &\leq rr_{h}^{RES} \Delta T_{h}^{RES} \label{eq:pl} \\
	&\forall h, &p_{G,h} &\leq \sum_{m} rr_{m}^{FRR} \Delta T_{h}^{RES} \label{eq:pg}
\end{align}

Substituting Eqs. \ref{eq:pl} and \ref{eq:pg} into Eq.~\ref{eq:deltaf} and using assumption \ref{as:ddf0} gives Eq. \ref{eq:FRRhm}.
\begin{align}\label{eq:FRRhm}
	&\forall h,\; \sum_{m} rr_{m}^{FRR} \Delta T_{h}^{RES} \geq  rr_{h}^{RES} \Delta T_{h}^{RES} + D\gls{df}
\end{align}

An algebraic constraint that allocates \gls{fcr} and \gls{frr} for short-term \gls{res} power variations compensation, is finally obtained based on ${D = \sum_{m} D_{m}}$ and that the \gls{df} range in steady state is $\pm r_{ss}$:
\begin{align}\label{eq:FRRhmsumD}
	&\forall h,\; \sum_{m} rr_{m}^{FRR} \Delta T_{h}^{RES} - \sum_{m} D_{m}r_{ss} \geq rr_{h}^{RES} \Delta T_{h}^{RES}
\end{align}

Note that the constraint in Eq. \ref{eq:FRRhmsumD} implies that \gls{fcr} and \gls{frr} are, when ${rr_{h}^{RES} > \sum_{m} rr_{m}^{FRR}}$, activated concurrently.
It is therefore implicitly assumed that sudden load variations and simultaneous worst-case \gls{res} power variations were taken into consideration when determining $P_{b,h}$ in \gls{fcr} sizing.
This leads to additional constraints, i.e. that generators $m$ participating in \gls{frr} have a total available up and down capacity of $P_{b,h}$ at each time step time $h$:
\begin{align}
    &\forall h, &\sum_{m} \left( P_{m}^{max} - P_{m,h} \right) &\geq P_{b,h} \label{eq:FRRup} \\
    &\forall h, &\sum_{m} \left( P_{m,h} - P_{m}^{min} \right) &\geq P_{b,h} \label{eq:FRRdown}
\end{align}

$P_{b,h}$ may eventually become asymmetrical, the positive worst-case power variation being different from the negative worst-case power variation.
The up and down-ramping capacities of generators can also be distinct, \gls{fcr} and \gls{frr} being asymmetrical in this situation.
Separate parameter sets $r_{ss}, r_{tr}, P_{b,h}, D_{m}, rr_{h,m}^{FRR}, rr_{h}^{RES}, \Delta T_{h}^{RES}$ should then be considered for up and down reserves.
The formulations proposed in this and the previous section can be developed further for this general case but will not, in the interests of brevity, be presented in this paper.

Regarding Assumption \ref{as:ddf0}, note that short-term power variations of renewables typically occur in ramp, as described later in \cref{sec:sizereserves}, and detailed in \ref{sec:scenario} and \cite{polleuxRelationshipBatteryPower2021} for solar \gls{pv} and \cite{hannesdottirDetectionCharacterizationExtreme2019} for wind power.
A cloud passage would likely take many seconds to complete shade the total area of a MW-scale plant, meaning that the power output would drop approximately in a ramp.
Where such ramps are smooth enough, the contribution of the \glspl{gt} inertia for the power balance represented in \cref{eq:deltaf} would indeed be minimal, as turbine governors would have enough time to react to the grid frequency changes and compensate the power imbalance. This is exactly what is being assumed in this formulation.

\begin{figure}
\centering
\includegraphics[width=1\linewidth]{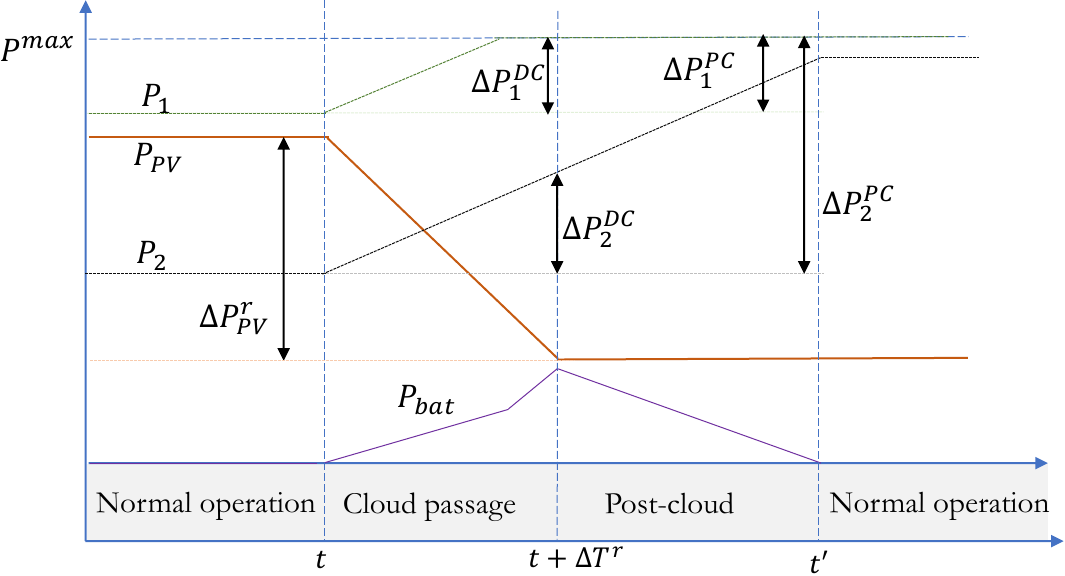}
\caption{Idealized response of an \glsxtrlong{aps} to a cloud-passage event.}
\label{fig:cloud-passage}
\end{figure}

This concept is better understood in \cref{fig:cloud-passage}, which sketches what happens when a cloud passage occurs in an \gls{aps} and where $P_1, P_2, P_{max}$ represent respectively the power delivered by two \glspl{gt} and their maximum limit, $P_{PV}$ and $P_{bat}$ are the power delivered by a solar \gls{pv} farm and a battery \gls{ess}, and the superscripts $DC$ and $PC$ indicates respectively the power variations during and post cloud passage.

The dynamics of short-term power variations are, in summary, different from sudden load or generation loss, the latter being addressed by the constraints proposed in \cref{sec:form1}.
The \glspl{gt} inertia will, in the latter case, be very important to guarantee the power balance represented in \cref{eq:deltaf} and avoid extreme frequency variations, a topic well explored in \cite{machowskiPowerSystemDynamics2008}, for instance.

\section{Case study of an industrial installation}
\label{sec:case}

\begin{figure}
\centering
\includegraphics[width=1\linewidth]{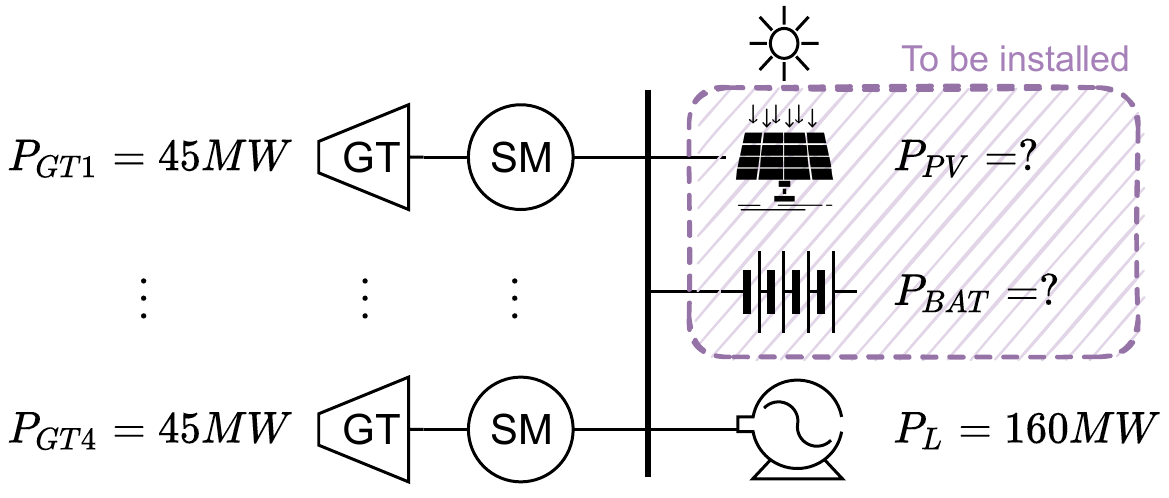}
\caption{Overview of the case study installation}
\label{fig:case-study}
\end{figure}

 The case study of an isolated \gls{og} installation with a peak electric power demand of \SI{160}{\mega\watt} is presented in this section, to exemplify the use of the proposed formulations.
This plant is currently equipped with four \glspl{gt} each of \SI{45}{\mega\watt}.
\SI{30}{\mega\watt} of the total load is considered to be non-essential, and can therefore be shed during an extreme contingency.
This design therefore allows the critical power demand of \SI{130}{\mega\watt} to be supplied, and a n+1 redundancy for \glspl{gt}.
The operator would like to integrate a solar \gls{pv} farm and a battery \gls{ess} into this plant, to reduce emissions of nitrogen oxides and \gls{ghg}.
\gls{fcr} and \gls{frr} should be supplied by the existing \glspl{gt}, the battery \gls{ess} being sized to provide only \gls{fcr}.
Figure \ref{fig:case-study} presents the proposed architecture and Table \ref{tab:case_study_params} the main technical parameters.

\begin{table}[htb]
\centering
\begin{tabular}{llr}
{Parameter} & {Unit} & {Value} \\ \hline
Rated frequency & \si{\hertz} & 50\\
$r_{ss}$ & \si{\hertz}  & 0.5 \\
Load & & \\
- Peak & \si{\mega\watt}  & 160 \\
- Critical & \si{\mega\watt}  & 130 \\
- Non-essential & \si{\mega\watt}  & 30 \\
\glspl{gt} & & \\
- Number & ~ & 4 \\
- Rated power & \si{\mega\watt} & 45 \\
- Droop & \si{\percent} & 10 \\
- Inertia & \si{\second}  & 5.51 \\
\end{tabular}
\caption{Case study parameters}
\label{tab:case_study_params}
\end{table}

\subsection{MILP algorithm formulation}\label{sec:milpalg}

A \gls{milp} algorithm was developed to support the operator's investment decision. 
The objective of optimization was to find the best performing pair of solar \gls{pv} (installed capacity $P_{PV}^{inst}$) and battery \gls{ess} (installed capacity $P_{bat}^{inst}$) for the plant.
Eq. \ref{eq:objective} shows the objective function, $c_{PV}, c_{bat}$ being the installation costs for solar \gls{pv} and battery \gls{ess} in \si{\$\per\kilo\watt}, $e$ denoting the discount rate of the project and
$c_f$ aggregated fuel costs including $CO_2$ and $NO_x$ penalties.
Operational costs across the plant's lifetime $Y_{inst}$ in years take into consideration the fuel consumption $FC_{m,h}$ of each generator $m$ at each time step $h$, as defined in Eq. \ref{cst_fuel_interp_y}.

\begin{align}
    &\min \; P_{PV}^{inst} c_{PV} + P_{bat}^{inst} c_{bat} + \sum_{y=0}^{Y_{inst}}  \sum_{h=1}^{8760} \sum_{m} \frac{{FC_{m,h} \, c_f}}{{(1+e)}^y} \label{eq:objective} \\
    &\forall m, \; \forall h, \; FC_{m,h} = a_m P_{m,h} + b_m \label{cst_fuel_interp_y}
\end{align}

Eqs.\ref{cst_load_balance}-\ref{cst_min_down_time} express the system operational constraints.
Eq. \ref{cst_load_balance} ensures that system power is in balance at each time step $h$.
Note that the \gls{ess} does not take part in permanent load balancing as it is used only for \gls{fcr}, that is ensuring frequency stability in case of short-term \gls{pv} drop or large contingency.
The energy flows related to charge and discharge therefore are considered negligible compared to the industrial load.
The injected solar \gls{pv} power $P_{PV,h}^{inj}$ is calculated in Eq. \ref{cst_PV_prod} using the available \gls{pv} area $A_{PV,h}$, the average irradiance $I_{h}$ in time step $h$, and derating factor $d_{PV}$.
Eq. \ref{cst_PV_curt} integrates $P_{PV}^{inst}$ into the objective function of Eq. \ref{eq:objective}.
\begin{align}
    &\forall h, &\sum_{m} P_{m,h}  &\geq P_{L,h} - P_{PV,h}^{inj}
\label{cst_load_balance} \\
    &\forall h, &P_{PV,h}^{inj} &\leq d_{pv} \, I_{h}\, A_{PV,h} \label{cst_PV_prod} \\
	&\forall h, &P_{PV}^{inst} &\geq I_{h}\, A_{PV,h} \label{cst_PV_curt}
\end{align}

Eqs. \ref{cst_fossil_fcr} to \ref{cst_min_down_time} express the operational constraints of each generator $m$ at each time step $h$,  $\rho_{m,h}$ representing the generator on/off status, $u_{m,h}$ and $ v_{m,h}$ start-up and shut down decisions, and $T_{m}^{up}$ and $T_{m}^{dn}$ minimum up and down-time.
\begin{align}
    &\forall m, \; \forall h, &P_{m,h}^{FCR} &\leq \rho_{m,h} D_{m} r_{ss}  \label{cst_fossil_fcr} \\
    &\forall m, \; \forall h, &P_{m,h} &\leq \rho_{m,h} P_m^{max} - P_{m,h}^{FCR} \label{cst_fossil_max} \\
    &\forall m, \; \forall h, &P_{m,h} &\geq \rho_{m,h} P_m^{min} + P_{m,h}^{FCR} \label{cst_fossil_min} \\
    &\forall m, \; \forall h, &u_{m,h} - v_{m,h} &\geq \rho_{m,h} - \rho_{m, h-1} \label{cst_status_change1} \\
    &\forall m, \; \forall h, &u_{m,h} + v_{m,h} &\leq 1 \label{cst_status_change_2} \\
    &\forall m, h \geq T_{m}^{up}, &\sum_{k=h-T_{m}^{up}}^{h-1} \rho_{m,k} &\geq T_{m}^{up} v_{m,h} \label{cst_min_ up_time} \\
    &\forall m, h \geq T_{m}^{dn}, &\sum_{k=h-T_{m}^{dn}}^{h-1} \rho_{m,k} &\leq T_{m}^{dn} ( 1 - u_{m,h}) \label{cst_min_down_time}
\end{align}

Note that Eqs.\ref{cst_fossil_max} and \ref{cst_fossil_min} allocate the \gls{fcr} for each generator $m$ defined in Eqs. \ref{eq:PmhFCR} and \ref{eq:Pmhdroop} according to a predefined damping $D_{m}$, calculated based on Table \ref{tab:case_study_params} data.
$D_{m} = P_m^{max} / (\beta_{m} P_{base})$.
$\beta_{m}$ is therefore the \gls{gt} droop, $P_{base}=\SI{45}{\mega\watt}$ is an arbitrary value adopted for normalization.
The total damping of the \glspl{gt} $D_{GT} = \sum_{m} D_{m}$ may not, however, be sufficient to comply with Eq. \ref{eq:sumDDmin}, battery \gls{ess} power $P_{bat}^{inst}$ in this case being sized to provide the remaining \gls{fcr}.
The constraints necessary for this are discussed in the next section, which are based on the formulations proposed in sections \ref{sec:form1} and \ref{sec:form2}.

\subsubsection{Battery sizing}\label{sec:sizereserves}
The first step in sizing spinning reserves is to apply sudden load or generation loss constraints, a simplified security assessment in this study being used.
The loss of a \gls{gt} in operation was therefore considered to be the worst contingency of this type. 
The system was therefore designed for n+1 redundancy of the \glspl{gt}, the sudden power disturbance term $P_{b,h}^{sud}$ therefore being defined by Eqs. \ref {eq:Pbhsudmax} and \ref{eq:Pbhsudmin}.
\begin{align}
 	&\forall m, \; \forall h, &P_{b,h}^{sud} &\leq P_m^{max}
 	\label{eq:Pbhsudmax} \\ 	
 	&\forall m, \; \forall h, &P_{b,h}^{sud} &\geq P_{m,h}
 	\label{eq:Pbhsudmin}
\end{align}

Cloud passage events at the solar \gls{pv} farm should also be taken into consideration.
Integration of these into the \gls{milp} formulation is however, in this study, based on the assumption that the optimization algorithm has access to a compact set $\textbf{H}^{max}$ of worst-case solar irradiance ramp events.
\ref{sec:scenario} summarizes a procedure for building $\textbf{H}^{max}$ from high-resolution irradiance time series.
Solar irradiance data provided by NREL \cite{sengupta_oahu_2010} was used.
The wavelet variability model \cite{laveWaveletBasedVariabilityModel2013} was also employed to take into account the geographical smoothing effect, using Sandia's PV\_LIB Toolbox for Matlab \cite{sandia_university_pv_2012}.
The datasets used to build $\textbf{H}^{max}$ in this study are available in \cite{alvesEfantnuDrep2021collabMINESParisTechNTNUV02022}, being $\textbf{H}^{max}$ for $h=11$ presented in Table \ref{tab:hull_drops}.
The duration and irradiance drop associated with ramp event $r$ in time step $h$ are denoted $\Delta T_{h,r}$ and $\Delta I_{h,r}^{PV}$.

\begin{table}
\centering
\begin{tabular}{c|rrrr}
\multicolumn{1}{c|}{Ramp} & \multicolumn{1}{c}{Durat.} & \multicolumn{1}{c}{Irr. drop}  \\
\multicolumn{1}{c|}{r}    &  \multicolumn{1}{c}{$\Delta T_{h,r}$} & \multicolumn{1}{c}{$\Delta I_{h,r}$} \\
\multicolumn{1}{c|}{~} & \multicolumn{1}{c}{[\si{\second}]} & [\si{\kilo\watt\per\meter\squared}]\\ \hline
$r_1$   & 2     &  0.061     \\
$r_2$   & 19    &  0.613      \\
$r_3$   & 36    &  0.778      \\
$r_4$   & 48    &  0.878      
\end{tabular}
\caption{The first 4 elements of the convex hull $\mathcal{H}^{max}_{11}$}
\label{tab:hull_drops}
\end{table}

The worst-case solar \gls{pv} disturbance term $P_{b,h}^{PV}$ at each time step $h$ can, where this framework is assumed, be specified using Eqs. \ref{eq:PbhPVmin} and \ref{eq:PbhPVmax}.
\begin{align}
 	&\forall h, \; \forall r, &P_{b,h}^{PV} &\geq d_{PV} \, \Delta I_{h,r} \, A_{PV,h} \label{eq:PbhPVmin} \\
 	&\forall h, &P_{b,h}^{PV} &\leq P_{PV,h}^{inj}
 	\label{eq:PbhPVmax}
\end{align}

Spinning reserves can be specified where both $P_{b,h}^{sud}$ and $P_{b,h}^{PV}$ have been defined and the formulations proposed in sections \ref{sec:form1} and \ref{sec:form2} are applied.
\Gls{frr} is sized using Eqs. \ref{eq:FRRupcs} and \ref{eq:FRRdowncs}, which are obtained when the constraints in Eqs. \ref{eq:FRRup} and \ref{eq:FRRdown} are used.

\begin{align}
    &\forall h, &\sum_{m} \left( \rho_{m,h} P_{m}^{max} - P_{m,h} \right) &\geq P_{b,h}^{sud} + P_{b,h}^{PV} \label{eq:FRRupcs} \\
    &\forall h, &\sum_{m} \left( P_{m,h} - \rho_{m,h} P_{m}^{min} \right) &\geq P_{b,h}^{sud} + P_{b,h}^{PV} \label{eq:FRRdowncs}
\end{align}

\Gls{fcr} is then calculated by combining the minimum damping requirements defined in Eq. \ref{eq:sumDDmin} and the constraint in Eq. \ref{eq:FRRhmsumD},
which assumes that $\Delta P_{h,r}^{PV}$ is partially compensated for by \gls{frr} during a cloud passage.
The battery \gls{ess} is therefore sized, at each time step $h$, to simultaneously compensate for any power deficit between a) $P_{b,h}^{sud}$, \gls{fcr} provided by \glspl{gt}; and b) $\Delta P_{h,r}^{PV}$, \gls{frr} provided by \glspl{gt} for all ramp events in $\textbf{H}^{max}$.
Eqs. \ref{eq:PbathFCRmax} and \ref{eq:PbathFCRmin} describe these requirements.
Finally, $P_{bat}^{inst}$ is integrated into the objective function of Eq. \ref{eq:objective} via Eq. \ref{cst_total_storage}.
Note that the proposed sizing is only addressing the maximum power of the \gls{ess}, not its energy capacity.
To calculate the energy flows demanded by \gls{fcr}, it is necessary to simulate the grid using more detailed models at sub-second time steps, which cannot be performed using the methodology here proposed.
This limitation is later exemplified and further discussed in \cref{sec:discussion}. 
\begin{align}
 	&\forall h, \forall r, &P_{bat,h}^{FCR} &\geq P_{b,h}^{sud} - P_{m,h}^{FCR} \nonumber \\
 	& & & + \Delta P_{h,r}^{PV} - \sum_{m} \left(\rho_{m,h}  rr_{m}^{FRR} \Delta T_{h,r} \right) \label{eq:PbathFCRmax} \\
 	&\forall h, &P_{bat,h}^{FCR} &\geq 0 \label{eq:PbathFCRmin} \\
    &\forall h, &P_{bat}^{inst} &\geq P_{bat,h}^{FCR} \label{cst_total_storage}
\end{align}

\subsection{Optimal sizing with static and dynamic frequency constraints}
\label{sec:case-comp}
\begin{table}
\centering
\begin{tabular}{llr}
{Parameter} & {Unit} & {Value} \\ \hline
$Y^{inst}$ & year & 20 \\
$rr_{m}$ & \si{\mega\watt\per\second} & 0.208 \\
$T_{m}^{up}$ & \si{\hour} &  6\\
$T_{m}^{dn}$ & \si{\hour}  & 6 \\
$a$ & \si{\cubic\meter\per\hour\per\kilo\watt} & 13782 \\
$b$ & \si{\cubic\meter\per\hour} & 5523 \\
$c_{fuel}$ & \si{\$\per\cubic\meter} & 1.01 \\
$c_{CO2}$ & \si{\$\per\tonne} & 120 \\
$ c_{PV}^{inst}$ & \si{\$\per\kilo\watt} & 400\\
$ d_{PV}$ & \si{\percent} & 80 \\
$ c_{bat}^{inst}$ & \si{\$\per\kilo\watt} & 250\\
$e$ & \si{\percent} & 3 
\end{tabular}
\caption{Techno-economic input parameters}
\label{tab:eco_params}
\end{table}

The \gls{milp} algorithm described in section \ref{sec:milpalg} was implemented in Gurobi 9.1, using an optimality gap tolerance of 1\%.
Table \ref{tab:eco_params} lists the techno-economic parameters used in the optimization.
Four scenarios were considered, these being selected to highlight the importance of frequency stability constraints:
\begin{enumerate}
    \item \textit{Baseline} is the current operation of the case study plant, power generation being based on 4 \glspl{gt}.
    \item \textit{No FC} includes the integration of the solar \gls{pv} farm, but without any frequency stability conditions. The constraints in section \ref{sec:sizereserves} are therefore omitted.
    \item \textit{Static FC} is the full implementation given in section \ref{sec:milpalg}. It adopts $P_{m,h}^{FCR} = 0$ in Eq. \ref{eq:PbathFCRmax}, which means that the frequency dynamics and the \gls{fcr} contribution during cloud passage are ignored. 
    \item \textit{Dynamic FC} includes the \gls{fcr} contribution in Eq. \ref{eq:PbathFCRmax}.
\end{enumerate}

\begin{table}
\centering
\begin{tabular}{rrrrr}
Indicator & Basel. & No FC & St. FC & Dyn. FC  \\ \hline
$P_{PV}^{inst}$  [\si{\mega\watt}]   & 0.0   & 129.76  & 62.00  & 62.00 \\
$P_{bat}^{inst}$ [\si{\mega\watt}]   & 0.0   & 0.00    & 33.3  & 10.80 \\
\glsentryshort{capex} [\si{\mega\$}] & 0.0   & 51.9  & 33.1 & 27.5   \\
$CO_2$ [\si{\mega\tonne\per year}]   & 111.3 & 95.4  & 105.7 & 105.7 \\
\glsentryshort{lcoe} [\si{\$\per\mega\watt\hour}] & 521.1 & 450.3 & 496.9 & 496.5 \\
Total costs [\si{\mega\$}]           & 9740   & 8870    & 9520   & 9580
\end{tabular}
\caption{Techno-economic optimization results for each scenario}
\label{tab:optim_res}
\end{table}

The results are given in Table \ref{tab:optim_res}.
The highest total costs are in the \textit{Baseline} case, minimum costs being in the \textit{No FC} case at a discount of \SI{8.9}{\percent} on \textit{Baseline}.
This can be explained by the installed capacity of the largest solar \gls{pv} being \SI{129.76}{\mega\watt} and the decrease of $CO_2$ emissions being \SI{14.3}{\percent} of \textit{Baseline}.
These factors also lead to the lowest \gls{lcoe} and highest \gls{capex} for all cases.
The \textit{No FC} case is, however, unrealistic, as it ignores frequency stability constraints and the need for energy storage. 
It does, however, provide useful references for maximum theoretical reductions in $CO_2$ emissions and \gls{lcoe}.

Adding spinning reserve constraints limits the potential of \gls{pv} integration, and requires battery \gls{ess} capacity,
this influencing the \gls{capex} in opposite directions.
The reduction of \gls{pv} installation costs is, however, bolder for the parameters of this case study.
\gls{capex} is reduced by \SI{36.2}{\percent} in the \textit{Static FC} and by \SI{47.0}{\percent} in the \textit{Dynamic FC} case, in relation to the \textit{No FC} case.
No significant differences are observed in fuel consumption, and therefore $CO_2$ emissions, between the two frequency constraints cases.
Their \gls{lcoe} and total costs therefore remain similar.
 \gls{lcoe} is reduced by \SI{13.5}{\percent} in \textit{No FC}, \SI{4.6}{\percent} in \textit{Static FC} and \SI{4.7}{\percent} in \textit{Dynamic FC} (on \textit{Baseline}).

It can be argued that the difference between \textit{Static FC} and \textit{Dynamic FC} is marginal for most indicators.
Note, however, that the battery \gls{ess} capacity in \textit{Dynamic FC} is reduced by \SI{67.6}{\percent} in relation to \textit{Static FC}.
This shows that \gls{ess} capacity can be reduced significantly where the tolerated frequency dynamics and complementary contribution of \gls{fcr} and \gls{frr} modeled in \cref{eq:PbathFCRmax} are taken into consideration during a cloud passage event.

\begin{figure}
\centering
\includegraphics[width=1\linewidth]{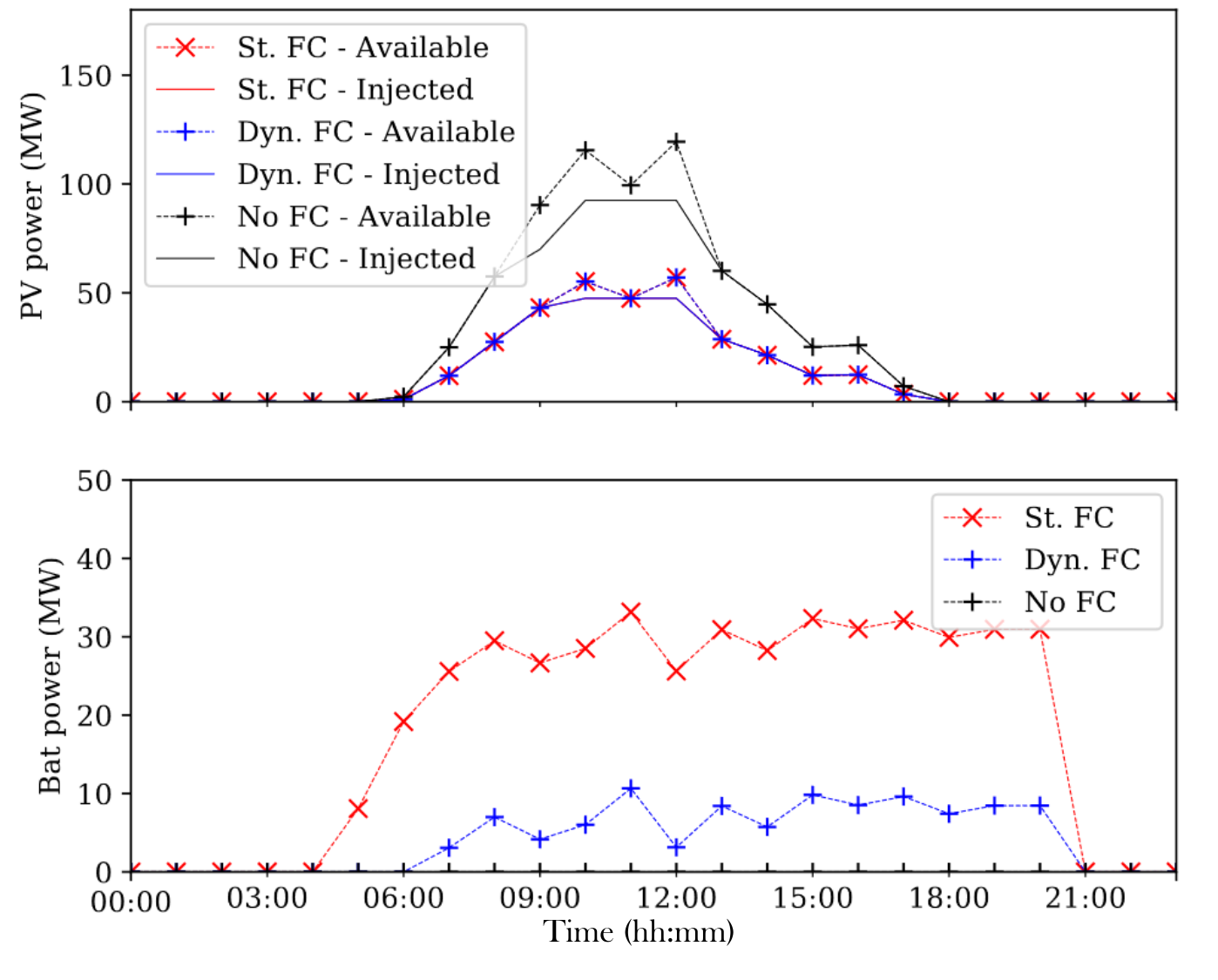}
\caption{Hourly available and injected PV profile and battery requirements on Jan 1st}
\label{fig:optim_dispatch}
\end{figure}

Figure \ref{fig:optim_dispatch} shows the hourly profiles on Jan 1st of available \gls{pv} power $P_{PV,h}^{avail} = I_{h} A_{PV,h}$, injected \gls{pv} power $P_{PV,h}^{inj}$, and the calculated battery contribution during a worst-case \gls{pv} power drop $P_{bat,h}^{FCR}$, and exemplifies what takes place at a more granular level.
Injected \gls{pv} power reaches \SI{92.6}{\mega\watt} at 11:00 in the \textit{No FC} case.
The injection is, however, curtailed to around \SI{54}{\mega\watt}in the {Static FC} and {Dynamic FC} cases,
the difference between these cases being explained by frequency stability constraints being neglected.
This allows the \gls{milp} algorithm to turn off one \gls{gt} and creates the opportunity for additional \gls{pv} capacity. 

\begin{table}
\centering
\begin{tabular}{c|ccc}
\multicolumn{1}{c|}{Ramp} & \multicolumn{1}{c}{St. FC} & \multicolumn{1}{c}{Dyn. FC} & \multicolumn{1}{c}{No FC} \\
\multicolumn{1}{c|}{r}    & \multicolumn{1}{c}{$P_{bat,r}^{FCR}$} & \multicolumn{1}{c}{$P_{bat,r}^{FCR}$} & \multicolumn{1}{c}{$P_{bat,r}^{FCR}$}   \\
\multicolumn{1}{c|}{~} & \multicolumn{1}{c}{[\si{\mega\watt}]} & \multicolumn{1}{c}{[\si{\mega\watt}]} & \multicolumn{1}{c}{[\si{\mega\watt}]} \\ \hline
$r_1$ & 25.7 & 3.2  & 11.3  \\
$\mathbf{r_2}$ & \textbf{33.3} & \textbf{10.8} & \textbf{50.2} \\
$r_3$ & 23.2 & 0.7 & 50.2 \\
$r_4$ & 15.3 & 0 & 48.1 
\end{tabular}
\caption{Required battery \gls{ess} capacity for the first 4 elements of the convex hull $\mathcal{H}^{max}_{11}$}
\label{tab:hull_results}
\end{table}

The infeasibility of the \textit{No FC} solution is shown by Table \ref{tab:hull_results},
the table specifying the power deficit between \gls{fcr}, \gls{frr}, $P_{b,h}^{sud}$ and $P_{b,h}^{PV}$ given by Eq. \ref{eq:PbathFCRmax} for the \gls{pv} ramps stated in Table \ref{tab:hull_drops}.
Remember that it is assumed that $P_{m,h}^{FCR} = 0$ in \textit{No FC} and \textit{Static FC}, in Eq. \ref{eq:PbathFCRmax}.
This assumption is also used in Eqs. \ref{cst_fossil_max} and \ref{cst_fossil_min} in \textit{No FC}.
The system may, in \textit{No FC}, face a power deficit of \SI{50.2}{\mega\watt}.
Critical \gls{pv} ramps are $r_2$ and $r_3$, which would undoubtedly lead to a grid blackout.
$P_{bat,r}^{FCR}$ is obtained, in the \textit{Static FC} case, by adding \SI{22.5}{\mega\watt} of one \gls{gt} loss to \SI{10.8}{\mega\watt} for the imbalance between \gls{frr} and \gls{pv} ramp rate.
The generator loss is fully compensated for in the \textit{Dynamic FC} case, by the damping of the remaining \glspl{gt}.
This shows the importance of taking into consideration the damping capabilities of generators participating in \gls{fcr} and the tolerated frequency deviations.

\subsection{Validation with time-domain simulations}

Time-domain simulations implemented in Matlab/Simulink (2021b) were used to validate the battery sizing by the proposed \gls{milp} algorithm.
Fig. \ref{fig:dynamic_model} presents a schematic diagram of this simulation model. 
The dynamics of the electricity grid are reduced into active power flows, Eq. \ref{eq:deltaf} being used to calculate grid frequency deviation.
Battery and solar \gls{pv} models consist of ideal active power sources, battery \gls{ess} adopting a droop control strategy including saturation and only providing \gls{fcr}.
\Glspl{gt} and their governors are represented by a droop model that contain a first-order filter to represent the delay of the actuator, being a time constant of \SI{0.5}{\second} employed.
Ramp-rate saturation is added for the \gls{frr} component.
Other relevant parameter values are given in Tables \ref{tab:case_study_params}-\ref{tab:optim_res}.

\begin{figure}
\centering
\includegraphics[width=1\linewidth]{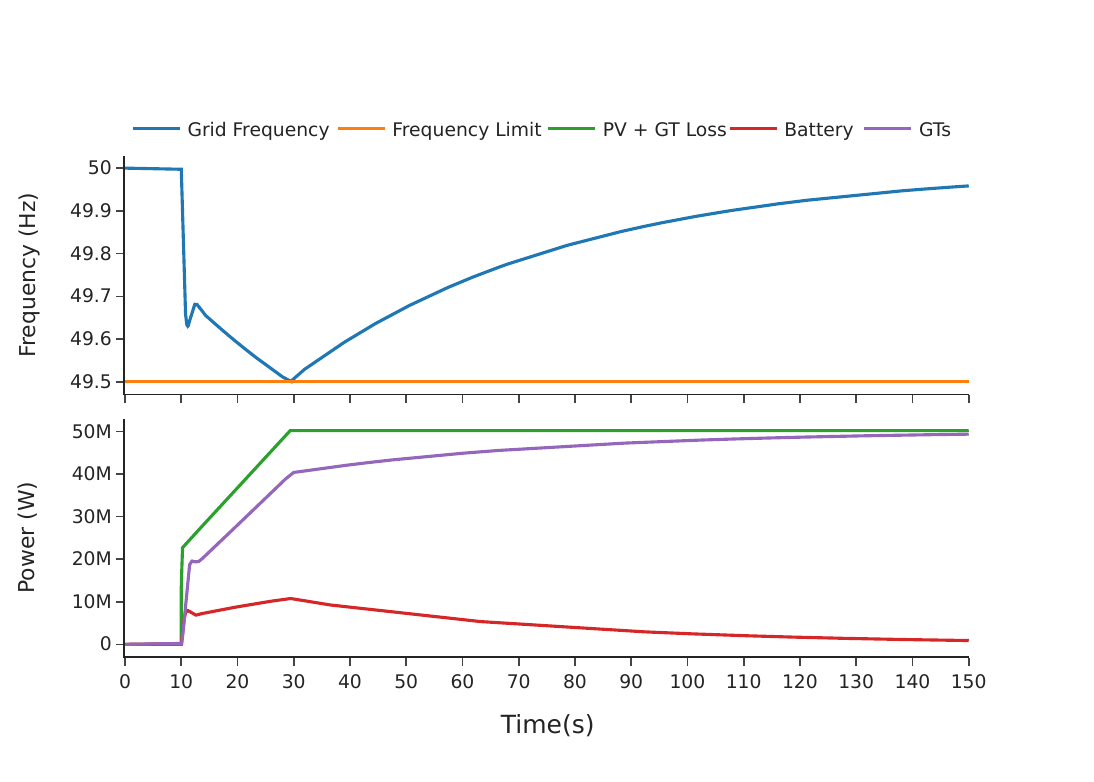}
\caption{Grid simulation of \gls{pv} ramp $r_2$ and one \gls{gt} loss at t=10sec}
\label{fig:ponctual_sim}
\end{figure}

The validation consists of simulating the worst-case cloud passage event and the simultaneous loss of one \gls{gt}, as presented in Table \ref{tab:hull_drops}, a check of whether frequency deviation is lower than or equal to $r_{ss}$ then being carried out.
Remark that all other cases will produce smaller frequency drops or lower \gls{frr} setpoint rate of change.
The power perturbation is generated by applying a load step that corresponds to $P_{b,h}^{sud}=\SI{22.5}{\mega\watt}$ at $t=\SI{10}{\second}$, and by simultaneously applying a negative power ramp corresponding to $r_2$.
The system inertia $M$ is equivalent to three \glspl{gt}.
Figure \ref{fig:ponctual_sim} shows the results of this simulation, including the battery \gls{ess}, the \glspl{gt} power and the grid frequency.
Battery support means the frequency drops to \SI{49.5}{\hertz} after the worst-case event, meeting the target of up to \SI{0.5}{Hz} deviation.
The grid frequency has also an oscillatory behavior after the sudden loss of one \gls{gt} due to the time delay of the turbine governors, which is compensated by the fast action of the battery \gls{ess}.
A further discussion of this topic is beyond the scope of this paper, but the interested reader can more information in \cite{motaOffshoreWindFarms2022}.

\section{Discussion}
\label{sec:discussion}

Optimal solar \gls{pv} and battery \gls{ess} capacities were calculated for an industrial plant, by using the formulation of linear frequency constraints and their integration in a \gls{milp} sizing problem.
Economic and environmental performance is reduced (at +10.8\% of $CO_2$ emissions and +8.0\% of total costs) in relation to optimization results without frequency constraints. 
It was, however, shown that the solution without frequency constraint is not resilient to worst-case events, such as a simultaneous  cloud passage and loss of a \gls{gt}.
Taking frequency reserves into consideration in sizing optimization therefore avoids the over-estimation of the benefits of solar power integration, and provides a more robust architecture in terms of power supply security.

The battery \gls{ess} capacity requirement was reduced by 67.6\% of the static calculation value that neglects the \gls{fcr} contribution.
This was achieved by taking into consideration the frequency deviation tolerance ($r_{ss}=0.5Hz$) in the \textit{Dynamic FC} scenario.
Considering simultaneous \gls{pv} drop and generator loss allows a robust approach, but also leads to conservative and costly \gls{ess} sizing.
The problem of battery \gls{ess} capacity allocation might, in less sensitive applications, take into consideration the highest value of the worst-case \gls{pv} ramp event and generator contingency, but not both simultaneously.
The impact of this on overall costs and $CO_2$ savings would, however, still be relatively small,
primarily due to the marginal effect that the proposed constraints would have on the fuel consumption of \glspl{gt}, which is the main cost driver in this case study.

The battery energy storage capacity in \si{\mega\watt\hour} for \gls{fcr} is not a dominant cost driver in this problem, and the cost function of \cref{eq:objective}, as consequence, do not consider it.
The actuation time of \gls{frr} is usually less than \SI{30}{\second} in a typical \gls{ems} for \gls{aps}, meaning that, if full \gls{fcr} capacity (i.e. \SI{10.8}{\mega\watt}) is delivered for 10 events every hour (i.e. 5 minutes in total), the required energy will be \SI{0.9}{\mega\watt\hour} for the selected battery \gls{ess} in the case study.
Typical losses of \glspl{ess} in the MW-range varies between 1 and 2\%, the required \gls{fcr} energy therefore being in the same order of magnitude of the battery \gls{ess} losses.
Remark that \gls{fcr} required energy represents only 0.56\% of the load demand in the case study (\SI{160}{\mega\watt\hour}) and can be assumed negligible when taking into consideration the accuracy of the model employed in the \gls{uc} problem and the optimality gap tolerance adopted.
Including the battery energy storage capacity in the problem formulation would however be extremely relevant if the \gls{ess} was designed to provide \gls{frr}, or if the proposed constraints were employed in an operational \gls{uc} problem designed to provide dispatch setpoints for the \glspl{gt} and \gls{ess} in real-time.

The attentive reader may have noticed that the fuel and $CO_2$ costs presented in Table \ref{tab:eco_params} are well above current market prices.
The reader may have also noticed that the project discount rate and installation costs for solar \gls{pv} and battery \gls{ess} are lower than the prevailing values in the industry.
This was to allow a high penetration of solar \gls{pv} in the case study installation, to highlight the impact of frequency stability constraints in this scenario.
The results of the \gls{milp} algorithm are, of course, sensitive to the economic parameters.
For example, if the discount rate is doubled, then the installed solar \gls{pv} capacity will drop to below \SI{9}{\mega\watt} in the \textit{Static FC} and \textit{Dynamic FC} scenarios.
The presentation of a sensitivity analysis in the case study is, however, beyond the scope of this work. 

Just considering Eqs. \ref{cst_fossil_max}, \ref{cst_fossil_min}, \ref{eq:FRRupcs} and \ref{eq:FRRdowncs} is, in terms of spinning reserves allocation, equivalent to the formulations proposed in \cite{schittekatte_impact_2018,rebollal_endogenous_2021}.
Introducing Eq. \ref{cst_total_storage}, however, contributes to security sizing problems, as it draws in both formulations.
\Gls{fcr} modeling during short-term power variations therefore improves the optimal solution.
The validation of the use of a time-domain model that only considers active power flows, shows that this frequency stability proposal can produce valid results. 
The power management of larger systems is, however, a multi-objective optimization problem with coupled variables, and one in which not only active power flows, but also voltages and reactive power flows, are optimized as discussed in \cite{mashayekh_integrated_2015}.
The inclusion of frequency constraints in such problems, to deal with short-term power variations, is however a topic for future research.

Note as well that a droop with a fixed value of 10\% for all \glspl{gt} was considered in the case study.
As briefly discussed in Section \ref{sec:milpalg}, the droop value directly affects total system damping.
It would therefore also affect the size of the battery \gls{ess}.
This observation suggests that this value should also be used as a decision variable in the optimization problem.
This does, however, lead to a non-linear formulation.
Further research on reserve allocation models and \gls{milp} formulation should therefore be carried out to circumvent this issue.

It is worth highlighting, at this point, that the constraints proposed in \cref{sec:form1,sec:form2} do not control $ \dot{\gls{df}} $, only assign the correct amount of \gls{fcr} given bounded values of $(\gls{pg} - \gls{pl})$ and $M$.
The proposed constraints, in other words, guarantee that the frequency deviation $\gls{df}$ remains within intervals $\pm r_{tr} $ during transient conditions and $\pm r_{ss} $ during steady-state conditions (i.e. post-disturbance), which is the definition of frequency stability given in \cite{hatziargyriouDefinitionClassificationPower2021} and in \cref{sec:freqstab}.

Note that large values of $\gls{df}$ in the negative direction (i.e. lack of power generation) can cause magnetic over-flux in electrical machines and transformers.
This situation not only increases losses and heating, but also induced voltage gradient between laminations that can break down the core insulation of these equipment and cause serious damages.
This topic is discussed in depth in \cite{alvesAnalysisOverexcitationRelaying2010,pyrhonenDesignRotatingElectrical2013,kulkarniTransformerEngineeringDesign2013,machowskiPowerSystemDynamics2008} and can be considered a more serious issue than large values of $\dot{\gls{df}}$ in low-inertia \gls{aps}.

Despite not being explicitly mentioned in \cref{sec:freqstab}, $M$ influences the term $r_{tr}$ in \cref{eq:dminrob}, which represents the maximum value of $\gls{df}$ during transient conditions, also known as frequency nadir or zenith.
This term can therefore be used to include constraints on $M$.
The model from \cref{eq:deltaf}, however, cannot capture the dynamics of the transient period, as it does not include the time delay of actuators.
References \cite{badesaSimultaneousSchedulingMultiple2019,trovatoUnitCommitmentInertiaDependent2019} have discussed this issue and proposed linearizations and/or analytical solutions that can be employed to include services such as \gls{ffr} or inertia emulation in the \gls{uc} problem.
The constraints proposed in these references can be included in the formulation proposed in \cref{sec:milpalg} where the dynamics of the transient period must be addressed.

In the case study case presented in \cref{sec:case}, however, a higher penetration of solar \gls{pv} is mainly prevented by the $\gls{df}$ constraints post-disturbance and the installation costs of the battery \gls{ess}, as shown in \cref{tab:optim_res} and discussed in \cref{sec:case-comp}.
The authors therefore consider reasonable to ignore the dynamics of the transient period in this specific case.
For a more general problem formulation, the constraints proposed in \cite{badesaSimultaneousSchedulingMultiple2019,trovatoUnitCommitmentInertiaDependent2019} can be combined with those presented in \cref{sec:form2}.
This however will turn the problem into a nonlinear optimization, as the ratio $\frac{D}{M}$ determines the $\gls{df}$ dynamics but each variable will be a decision variables in the optimization.
Addressing this issue and developing a convex reformulation of the problem that can be solved by \gls{milp} is a relevant topic for future work.

\section{Conclusion}
\label{sec:conclusion}

The problem of spinning reserves allocation for isolated grids with high penetration of \glspl{res} is addressed in this work.
Linear frequency stability constraints were first formulated, to allow integration in a frequency constrained \glsxtrfull{uc} problem, the proposed constraints ensuring the allocation of enough reserves to compensate for short-term renewable variations and generator contingencies. 
This includes practical issues such as the limited ramping capacities of \glsxtrfull{frr}.
It also takes advantage of short-term frequency variations and the complementary action between \glsxtrfull{fcr} and \glsxtrfull{frr}.

The reserve allocation strategy is exemplified by a case study in which a \glsxtrfull{milp} algorithm is formulated for the optimal sizing of a solar \glsxtrfull{pv} farm and an \glsxtrfull{ess}.
Linear frequency stability constraints are aggregated to ensure the resiliency of the grid, in the event of cloud passage and generator contingency occurring simultaneously.
The results show that this method provides an optimal and secure architecture.
It also shows that neglecting stability constraints leads to an inoperable solution, and overestimates system profitability and \glsxtrfull{ghg} emission savings, by 8.0 and 10.8\% respectively.

\appendix

\section{Generation of solar \gls{pv} power drop scenarios}
\label{sec:scenario} 

\begin{figure*}
\centering
\includegraphics[width=\linewidth]{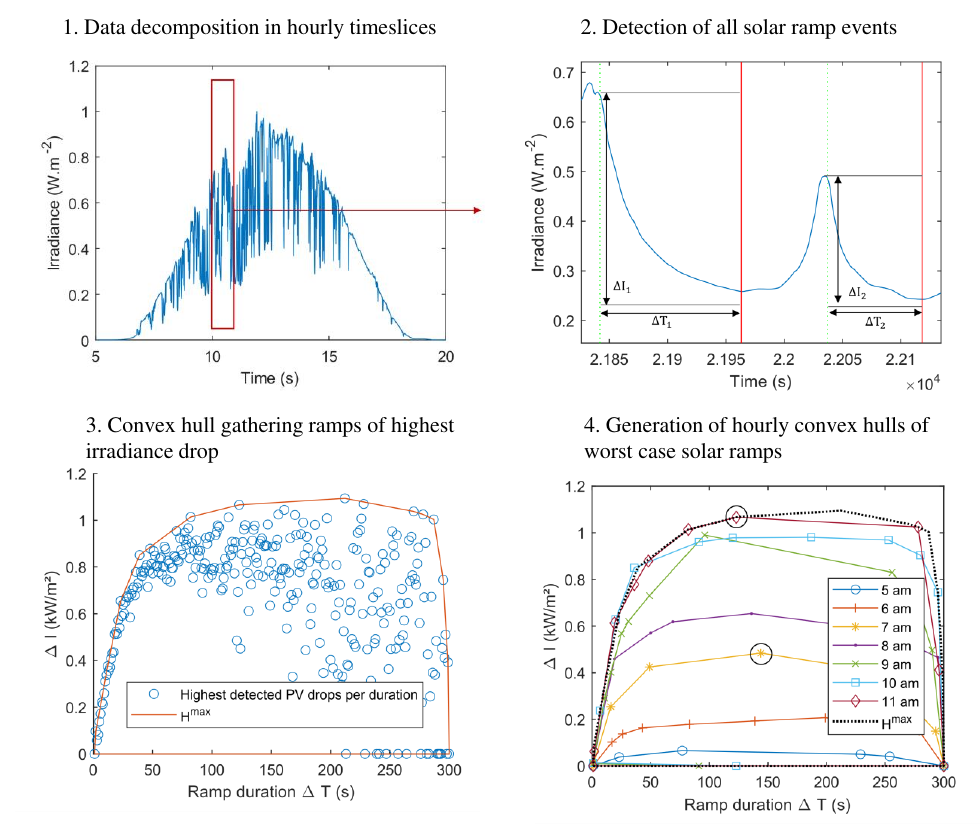}
\caption{Process of worst-case ramp identification from high-resolution timeseries to convex hulls $\mathcal{H}_h^{max}$ in the lower-right figure}
\label{fig:ramp_process}
\end{figure*}

Solar variability is typically measured by irradiance sensors, which generate high-resolution time series.
The measurement sampling time must be lower than \SI{5}{\second} if cloud passage events are to be captured.
Integrating these time series into a high-level optimization is however impractical, due to time increment discrepancies.
Fig. \ref{fig:ramp_process} summarizes a procedure for extracting ramp scenarios from irradiance sensor time series, as was recently proposed in \cite{polleuxRelationshipBatteryPower2021} and applied in section \ref{sec:case}.

In step 1, high resolution data from sensors is decomposed into hourly time slices.
In step 2, all ramp events within an hour slice are detected.
A ramp event $rr_{PV,h,r}$ is defined by its irradiance drop $\Delta I_{h,r}$ and ramp duration $\Delta T_{h,r}$ during cloud passage.
In step 3, all events are grouped in a compact set.
The hourly convex hull $\mathcal{H}_h^{max}$ is defined to gather the set of highest irradiance drops for each duration, to allow only the worst-case events to be extracted.
This provides a subset of a limited number of ramps.
Steps 2 and 3 are then repeated for each hour slice defined in step 1.
Finally, in step 4, all $\mathcal{H}_h^{max}$ are grouped and the global convex compact worst-case ramp set $\mathcal{H}^{max}$ is obtained.

The solar \gls{pv} power drop $\Delta P_{h,r}^{PV}$ is calculated using the irradiance drop according to Equation \ref{eq:pv_drop_curtailed}, $A_{PV,h}, I_{h}, d_{PV}$ being the available \gls{pv} area, the average irradiance, and derating factor, respectively.
\begin{equation}
	\Delta P_{h,r}^{PV} = d_{PV} \, \Delta I_{h,r} \, A_{PV,h}
\label{eq:pv_drop_curtailed}
\end{equation}

\pagebreak

\section{Power system model used for validation}
\label{sec:modelvalid} 

\begin{figure*}[h]
\centering
\includegraphics[width=0.8\linewidth]{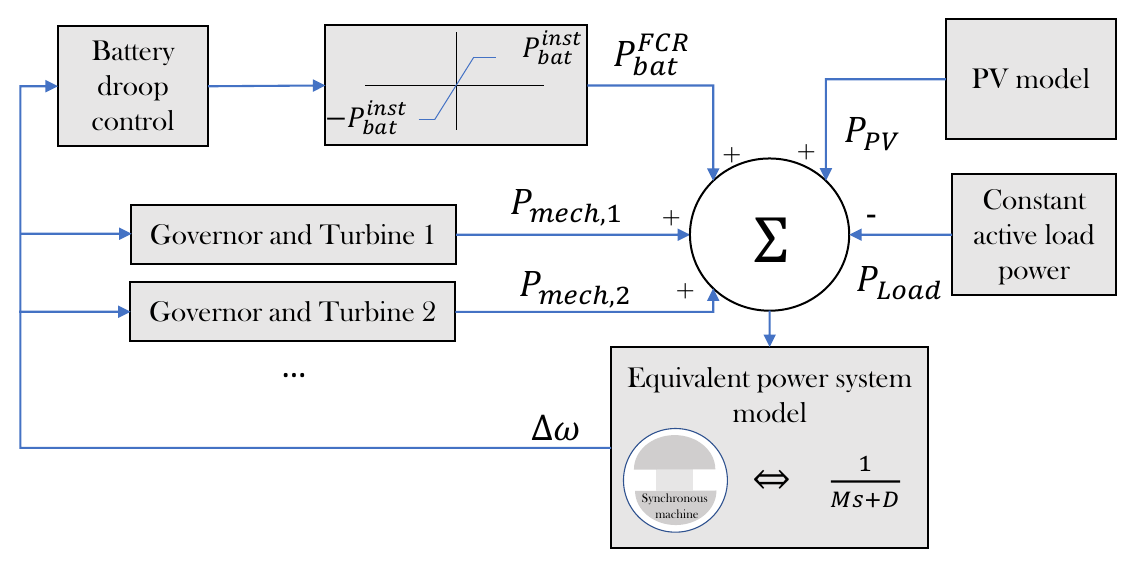}
\caption{Power system model implemented in Matlab/Simulink}
\label{fig:dynamic_model}
\end{figure*}

\pagebreak


 \bibliographystyle{elsarticle-num} 
 \bibliography{cas-refs}





\end{document}